\documentclass[epjST]{svjour}
\usepackage{bm}
\usepackage{amsmath}
\usepackage{amssymb}

\usepackage{graphics}
\usepackage{graphicx}

\DeclareMathOperator{\sgn}{sgn}

\DeclareMathOperator{\Rea}{\mathrm{Re}}
\DeclareMathOperator{\Sp}{\mathrm{Sp}}

\begin{document}
\title{Long-range Josephson effect controlled by temperature gradient and circuit topology}
\subtitle{}
\author{Mikhail S. Kalenkov\inst{1} \and Andrei D. Zaikin\inst{2}\fnmsep\inst{3}\fnmsep\thanks{\email{andrei.zaikin@kit.edu}}}%
\institute{
I.E. Tamm Department of Theoretical Physics, P.N. Lebedev Physical Institute, 119991 Moscow, Russia 
\and  
Institute for Quantum Materials and Technologies, Karlsruhe Institute of Technology (KIT), 76021 Karlsruhe, Germany  
\and 
National Research University Higher School of Economics, 101000 Moscow, Russia}
\abstract{
We demonstrate that the supercurrent can be strongly enhanced in cross-like superconducting hybrid nanostructures ($X$-junctions) exposed
to a temperature gradient. At temperatures $T$ exceeding the Thouless energy of our $X$-junction the Josephson current  decays algebraically with increasing $T$ and can be further enhanced by a proper choice of the circuit topology. At large values of the temperature gradient the non-equilibrium contribution to the supercurrent may become as large as the equilibrium one at low $T$. We also predict a variety of transitions between 0- and $\pi$-junction states controlled by the temperature gradient as well as by the system geometry. Our predictions can be directly verified in modern experiments.} 
\maketitle
\section{Introduction}
\label{intro}
Equilibrium Josephson current between two superconductors depends periodically on the phase difference $\chi$ between them \cite{Jos}. The magnitude of this current also depends on temperature $T$. This dependence varies in different types of superconducting weak links and is perhaps most strongly pronounced in superconducting junctions containing a sufficiently thick layer of a normal metal \cite{ZZh,GreKa,BWBSZ1999,GKI}. In these so-called SNS junctions the supercurrent reaches its maximum value $I_C (0) \simeq 10.82 E_{\mathrm{Th}}/(eR_n)$ \cite{GreKa} at $T \to 0$, whereas at higher temperatures $T> E_{\mathrm{Th}}$ this current reduces exponentially as $\propto e^{-\sqrt{2 \pi T/E_{\mathrm{Th}}}}$, where $E_{\mathrm{Th}}$ and $R_n$ are respectively an effective Thouless energy and a normal state resistance of an SNS device.

The supercurrent flowing across an SNS junction can be significantly affected by driving the electron distribution function out of equilibrium. Such non-equilibrium conditions can be achieved, e.g., by applying an external ac signal \cite{Aslamazov82,Zaikin83} or a dc voltage $V$ in the cross-like geometry considered in \cite{Volkov,WSZ,Yip,Teun}. In the first case one can observe a strong supercurrent stimulation at $T> E_{\mathrm{Th}}$, while in the second one by tuning $V$ one can realize the transition to a $\pi$-junction state.

Yet another way to drive the electron distribution function inside an SNS junction out of equilibrium is to expose it to a thermal gradient.
Recently it was demonstrated \cite{KDZ20} that by doing so one can effectively support long-range phase coherence of quasiparticles inside the N-layer at temperatures above the Thouless energy where the equilibrium supercurrent already becomes vanishingly small. Specifically, for the so-called $X$-junction geometry illustrated in Fig. 1 in the high temperature limit $T_{1,2}\gg E_{\mathrm{Th}}$  -- up to some geometry factors -- one finds $I_C \sim  I_C(0) E_{\mathrm{Th}}|1/T_1 - 1/T_2|$, where $T_1$ and $T_2$ are different temperatures at which 
two normal terminals N$_1$ and N$_2$ are maintained, see Fig. 1. Hence, this current turns out to be a lot bigger than the equilibrium one at any of the two temperatures $T_1$ or $T_2$. 

Furthermore, under such non-equilibrium conditions the junction is described by a non-sinusoidal current-phase relation and may exhibit a pronounced $\pi$-junction-like behavior. It was argued \cite{KDZ20} that all these non-trivial features are caused by the presence of non-equilibrium low energy quasiparticles suffering little dephasing while propagating across the N-layer. We also note that a somewhat similar situation was encountered for the Aharonov-Bohm effect in superconducting heterostructures containing a normal metallic loop, see, e.g., Refs. \cite{BWBSZ1999,Gre,GWZ}.

In this paper we will extend the work \cite{KDZ20} in several important aspects. In particular, here we will lift the symmetry restrictions adopted
in \cite{KDZ20} and evaluate the Josephson current across a general asymmetric $X$-junction exposed to an arbitrary temperature gradient.
We will demonstrate that, on one hand, the presence of electron-hole asymmetry weakly affects the Josephson current and, on the other hand, that the effect of supercurrent stimulation can be substantially enhanced by a proper choice of the circuit topology. 
\begin{figure}
\begin{center}
\includegraphics[width=70mm]{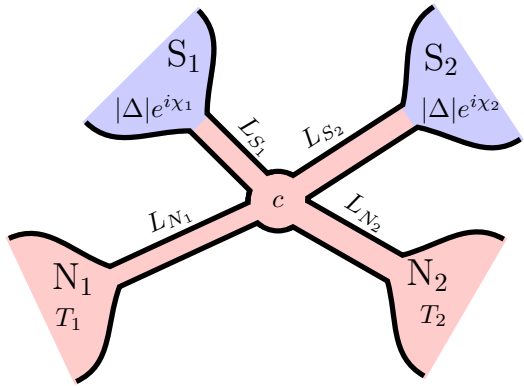}
\end{center}
\caption{$X$-junction structure under consideration.}
\label{ssXnn-fig}
\end{figure}

\section{The model and basic formalism}
Following Ref. \cite{KDZ20} we are going to consider an $X$-junction which consists of two superconducting and two normal terminals interconnected by four normal metallic wires of lengths $L_{S_{1,2}}$, $L_{N_{1,2}}$ and cross sections ${\mathcal A}_{S_{1,2}}$, ${\mathcal A}_{N_{1,2}}$ respectively, as shown in Fig. 1. The two normal terminals are disconnected from any external circuit and are maintained at different temperatures $T_1$ and $T_2$. The supercurrent $I_S(\chi)$ can flow between superconducting terminals S$_1$ and S$_2$ with the phase difference $\chi=\chi_1-\chi_2$.

In what follows we will employ the standard quasiclassical formalism based on the Usadel equations  combined with the Keldysh technique \cite{BWBSZ1999}
\begin{equation}
i D \nabla \left(\check G  \nabla \check G\right)=
\left[\check \Omega, \check G \right], \quad \check G \check G=\check 1,
\label{Usadel}
\end{equation}
where $D$ is the diffusion constant, $\check \Omega$ is $4\times4$ matrix
\begin{equation}
\check \Omega
=
\begin{pmatrix}
\hat \Omega^R & 0 \\
0 & \hat \Omega^A
\end{pmatrix},
\quad
\hat \Omega^R = \hat \Omega^A
=
\begin{pmatrix}
\varepsilon + eV & \Delta \\
-\Delta^* & - \varepsilon + eV
\end{pmatrix},
\end{equation}
where $\varepsilon$, $V$ and $\Delta$ denote respectively quasiparticle energy, electrostatic potential and superconducting order parameter. Equations (\ref{Usadel}) allow to evaluate the quasiclassical Green function $\check G$, which is represented by the $4\times4$ matrix in the Keldysh$\otimes$Nambu space
\begin{equation}
\check G = 
\begin{pmatrix}
\hat G^R & \hat G^K \\
0 & \hat G^A
\end{pmatrix},
\quad
\hat G^{R,A} =
\begin{pmatrix}
G^{R,A} & F^{R,A} \\
\tilde F^{R,A} & - G^{R,A}
\end{pmatrix},
\end{equation}
where $\hat G^{R,A}$ are retarded and advanced Green functions. It is convenient to parametrize the Keldysh matrix function $\hat G^K$ as
\begin{equation}
\hat G^K = \hat G^R \hat h - \hat h \hat G^A,
\quad
\hat h = h^L + \hat \tau_3 h^T,
\end{equation}
where $h^L$ and $h^T$ are the parts of the quasiparticle distribution function. The current density can be calculated by means of standard relation
\begin{gather}
\bm{j} =  -\dfrac{\sigma}{8 e }
\int
d \varepsilon
\Sp (\hat \tau_3\check G \nabla \check G)^K.
\end{gather}

Inside the normal wires the functions $h^L$ and $h^T$ obey the equations 
\begin{gather}
iD\nabla\left[ D^T\nabla h^T + {\mathcal Y}\nabla h^L + \bm{j}_{\varepsilon} h^L \right]=0,
\label{Tkin}
\\
iD\nabla\left[ D^L\nabla h^L - {\mathcal Y} \nabla h^T + \bm{j}_{\varepsilon} h^T \right]=0,
\label{Lkin}
\end{gather}
which follow directly from Eqs. (\ref{Usadel}). Here 
\begin{gather}
D^T = \dfrac{1}{4}\Sp (1 - \hat \tau_3 \hat G^R \hat \tau_3 \hat G^A)
=\dfrac{1}{4}
\left[ 2 - 2 G^R G^A + F^R \tilde F^A + \tilde F^R F^A \right],
\\
D^L= \dfrac{1}{4}\Sp (1 - \hat G^R \hat G^A)
=\dfrac{1}{4}
\left[ 2 - 2 G^R G^A - F^R \tilde F^A - \tilde F^R F^A \right]
\end{gather}
denote dimensionless diffusion coefficients,  
\begin{equation}
\mathcal{Y}=
\dfrac{1}{4}
\left[ \tilde F^R F^A - F^R \tilde F^A   \right]
\end{equation}
is the kinetic coefficient which accounts for the presence of the particle-hole asymmetry in our system and
\begin{equation}
\bm{j}_{\varepsilon}=\dfrac{1}{2}\Rea\left(F^R \nabla \tilde F^R - \tilde F^R \nabla F^R \right)
\end{equation}
defines the spectral supercurrent.

\section{General analysis}
For quasi-one-dimensional geometry of the normal wires adopted here the kinetic equations \eqref{Tkin}, \eqref{Lkin} can be solved exactly. Let us first rewrite these equations in the matrix form
\begin{gather}
\hat D 
\begin{pmatrix}
h^T \\ h^L
\end{pmatrix}'
+
j_{\varepsilon}
\hat \tau_1
\begin{pmatrix}
h^T \\ h^L
\end{pmatrix}
=
-\dfrac{e}{\sigma \mathcal{A}}
\begin{pmatrix}
I^T\\I^L
\end{pmatrix},
\quad\hat D =
\begin{pmatrix}
D^T & \mathcal{Y} \\
- \mathcal{Y} & D^L
\end{pmatrix}
\label{kin2}
\end{gather}
where $\mathcal{A}$ is the cross section of the corresponding wire segment. The spectral currents $I^T$ and $I^L$ represent the integration constants independent of the coordinate along the wire. Below we will use the convention according to which the current is positive if it flows from the corresponding terminal to the crossing point $c$.

As the spectral supercurrent $j_{\varepsilon}$ vanishes identically in the wires attached to the normal terminals, Eq. 
\eqref{kin2} can be easily integrated, and we get
\begin{equation}
\hat G_{N_i}
\begin{pmatrix}
h^T_c - h^T_{N_{i}}
\\
h^L_c - h^L_{N_{i}}
\end{pmatrix}
=
\begin{pmatrix}
I^T_{N_i} \\ I^L_{N_i}
\end{pmatrix},
\quad
\hat G_{N_i}
=
\begin{pmatrix}
G^T_{N_i} & G^{\mathcal{Y}}_{N_i} \\
- G^{\mathcal{Y}}_{N_i} & G^L_{N_i}
\end{pmatrix}
=
\left[
\int_{L_{N_i}} 
\dfrac{\hat D^{-1} dx}{ \sigma \mathcal{A}_{N_i} }
\right]^{-1}.
\end{equation}
The spectral conductances $G^T_{N_i}$, $G^L_{N_i}$ and $G^{\mathcal{Y}}_{N_i}$ all exhibit a nontrivial energy dependence in the vicinity of the Thouless energy, whereas for $|\varepsilon | \gg E_{\mathrm{Th}}$ the conductance $G^{\mathcal{Y}}_{N_i}$ tends to zero and $G^T_{N_i}$, $G^L_{N_i}$ just reduce to their normal state values $G^n_{N_i}\equiv 1/R^n_{N_i}=\sigma \mathcal{A}_{N_i}/ L_{N_i}$. 

In the wires attached to the superconducting terminals and at subgap energies the spectral currents $I^L_{S_i}$ vanish identically. With this in mind, we obtain
\begin{gather}
G^T_{S_1} h_c^T + \mathcal{G} h_c^L = -e I^T_{S_1},
\quad
G^T_{S_2} h_c^T - \mathcal{G} h_c^L = -e I^T_{S_2},
\end{gather}
where
\begin{equation}
\mathcal{G} (\varepsilon )= \mathcal{A}_{S_1} \sigma j_{1\varepsilon} = - \mathcal{A}_{S_2} \sigma j_{2\varepsilon}.
\end{equation}
Here we also made use of the condition $h^T=0$ which holds for both SN interfaces at subgap energies. In general the spectral conductances $G^T_{S_{1,2}}$ depend on the kinetic coefficients $D^{T,L}$ and $\mathcal{Y}$ in a complicated manner. These conductances demonstrate a nontrivial energy dependence below the Thouless energy and tend to their normal state values $G^n_{S_{1,2}} =\sigma \mathcal{A}_{S_{1,2}}/ L_{S_{1,2}}$ in the high energy limit. 

The spectral current conservation conditions at the crossing point $c$ take the form
\begin{gather}
I^T_{S_1} + I^T_{S_2} + I^T_{N_1} + I^T_{N_2} =0,
\quad
I^L_{N_1} + I^L_{N_2} =0.
\end{gather}
Resolving the above equations we can express all the spectral currents $I^{T,L}$ in terms of the distribution functions in the normal terminals
\begin{equation}
h^{T/L}_{N_{1,2}}
=
\dfrac{1}{2}
\left[
\tanh \dfrac{\varepsilon + e V_{1,2}}{2 T_{1,2}}
\mp
\tanh \dfrac{\varepsilon - e V_{1,2}}{2 T_{1,2}}
\right].
\label{bou1}
\end{equation}
As soon as the spectral currents are established,  the electric currents $I_{X_i}$ in all four normal wires can be recovered with the aid of a simple formula
\begin{equation}
I_{X_i} = \dfrac{1}{2} \int I_{X_i}^T d \varepsilon,
\quad
X= S, N,
\quad
i=1,2.
\end{equation}

\section{Supercurrent in the presence of a temperature gradient}
Without any temperature gradient no voltage drop across the normal terminals N$_1$ and N$_2$ can occur. However, different electrostatic potentials $V_1$ and $V_2$ at these terminals are in general induced provided they are kept at different temperatures $T_1$ and $T_2$.  
This is a manifestation of the so-called thermoelectric effect \cite{Ginzburg}. In order to evaluate thermoelectric voltages $V_1$ and $V_2$ 
it is necessary to bear in mind that no currents can flow trough the normal terminals, i.e. $I_{N_1} = I_{N_2} =0$. Making use of these conditions,
in the limit $E_{\mathrm{Th}} \ll T_{1,2} \ll \Delta$ one arrives at the result  \cite{KDZ20}
\begin{multline}
e V_1=
-
\dfrac{1}{4}
\left(\dfrac{1}{T_1} - \dfrac{1}{T_2}\right)
\int
\dfrac{\varepsilon d \varepsilon}{G_{N_1}^n }
\Biggl\{
\dfrac{G_{N_1}^{\mathcal{Y}} G_{N_2}^L}{G_{N_1}^L + G_{N_2}^L} 
+
\dfrac{G_{N_2}^{\mathcal{Y}} G_{N_1}^L -G_{N_1}^{\mathcal{Y}} G_{N_2}^L}{
\det|\hat G_{S_1} + \hat G_{S_2} + \hat G_{N_1} + \hat G_{N_2}|}
\\\times
\Biggl[
G_{N_1}^T - G_{N_1}^n \dfrac{G_{S_1} + G_{S_2}}{G^n_{S_1} + G^n_{S_2}}
+
\dfrac{G_{N_1}^{\mathcal{Y}} ( G_{N_1}^{\mathcal{Y}} + G_{N_2}^{\mathcal{Y}} )}{G_{N_1}^L + G_{N_2}^L} 
\Biggr]
\Biggr\},
\label{V1}
\end{multline}
which allows to express $V_1$ in terms of the spectral conductances. The thermoelectric voltage $V_2$ is determined simply by interchanging the indices  $1\leftrightarrow2$ in Eq. \eqref{V1}. 

We observe that the voltages $V_{1,2}$ may differ from zero only provided the conductances $G_{N_{1,2}}^{\mathcal{Y}}$ do not vanish, which is the case in the presence of electron-hole asymmetry. It is straightforward to demonstrate \cite{KDZ20} that under extra symmetry conditions (i) $L_{S_1} = L_{S_2}$ and (ii) ${\mathcal A}_{S_1} = {\mathcal A}_{S_2}$ the kinetic coefficient ${\mathcal Y}$ equals to zero everywhere in the N-wires attached to normal terminals N$_1$ and N$_2$. Hence, no electron-hole asymmetry occurs in this case and $V_{1,2}\equiv 0$. The Josephson current across our $X$-junction was analyzed in Ref. \cite{KDZ20} only under the conditions (i) and (ii), i.e. in the absence of any induced thermoelectric potentials.

In this work we lift both symmetry conditions (i) and (ii)  and evaluate the supercurrent $I_S$ in the presence of electron-hole asymmetry and the thermoelectric effect. The magnitude of thermoelectric potentials $|V_{1,2}|$ may be not small in this case, in some special limits reaching the values of up to $\sim E_{\mathrm{Th}}/e$ \cite{KDZ20}. Hence, this effect should in general be included into our consideration.

Employing the quasiclassical formalism outlined in the previous sections, after some algebra we obtain
\begin{multline}
I_S
=
-
\dfrac{1}{4e}
\int
\dfrac{G_{S_1} - G_{S_2}}{\det|\hat G_{S_1} + \hat G_{S_2} + \hat G_{N_1} + \hat G_{N_2}|}
\Bigl\{
[(G^L_{N_1} + G^L_{N_2})G^T_{N_1} + (G^{\mathcal{Y}}_{N_1} + G^{\mathcal{Y}}_{N_2})G^{\mathcal{Y}}_{N_1}] h^T_{N_1} 
\\-
[G^L_{N_1}G^{\mathcal{Y}}_{N_2} - G^{\mathcal{Y}}_{N_1} G^L_{N_2}] (h^L_{N_1} - h^L_{N_2})
+
[(G^L_{N_1} + G^L_{N_2})G^T_{N_2} + (G^{\mathcal{Y}}_{N_1} + G^{\mathcal{Y}}_{N_2})G^{\mathcal{Y}}_{N_2}] h^T_{N_2} 
\Bigr\}
d \varepsilon
\\-
\dfrac{1}{4}
\int
\dfrac{2 \mathcal{G}}{\det|\hat G_{S_1} + \hat G_{S_2} + \hat G_{N_1} + \hat G_{N_2}|}
\Bigl\{
[G^{\mathcal{Y}}_{N_2}G^T_{N_1} - G^{\mathcal{Y}}_{N_1} (G^T_{N_2} + G_{S_1} + G_{S_2})] h^T_{N_1} 
\\+
[(G^T_{N_1} + G^T_{N_2} + G_{S_1} + G_{S_2})G^L_{N_1} + (G^{\mathcal{Y}}_{N_1} + G^{\mathcal{Y}}_{N_2}) G^{\mathcal{Y}}_{N_1}] h^L_{N_1}
+
[G^{\mathcal{Y}}_{N_1}G^T_{N_2} - G^{\mathcal{Y}}_{N_2} (G^T_{N_1} + G_{S_1} + G_{S_2})] h^T_{N_2} 
\\+
[(G^T_{N_1} + G^T_{N_2} + G_{S_1} + G_{S_2})G^L_{N_2} + (G^{\mathcal{Y}}_{N_1} + G^{\mathcal{Y}}_{N_2}) G^{\mathcal{Y}}_{N_2}] h^L_{N_2}
\Bigr\}
d \varepsilon.
\end{multline}
This cumbersome expression can be cast to the form
\begin{equation}
I_S
=
r_{N_2}^n I_J(T_1,\chi) + r_{N_1}^n I_J(T_2,\chi)
+ I_S^{\mathrm{ne}},
\label{IS}
\end{equation}
where $r_{N_{1,2}} = R^n_{N_{1,2}}/(R^n_{N_1} + R^n_{N_1})$ and
\begin{equation}
I_J(T,\chi)= - \dfrac{1}{2e}\int \mathcal{G} (\varepsilon )\tanh \dfrac{\varepsilon}{2T} d \varepsilon
\label{IJ}
\end{equation}
is the equilibrium Josephson current across our $X$-junction.  At sufficiently high temperatures $T \gg E_{\mathrm{Th}}$ Eq. (\ref{IJ})
yields
\begin{equation}
I_J= \dfrac{16 \varkappa}{ 3 + 2\sqrt{2}}\dfrac{E_{\mathrm{Th}}}{eR_n^a}\left(\dfrac{2 \pi T}{E_{\mathrm{Th}}}\right)^{3/2}
e^{-\sqrt{2 \pi T /E_{\mathrm{Th}}}} \sin\chi ,
\label{IJeq}
\end{equation}
where we defined  $\varkappa = 4 \sqrt{\mathcal{A}_{S_1}\mathcal{A}_{S_2}}/(\mathcal{A}_{S_1} + \mathcal{A}_{S_2} + \mathcal{A}_{N_1} + \mathcal{A}_{N_2})$, $R_n^a = L_S/ (\sigma  \sqrt{\mathcal{A}_{S_1} \mathcal{A}_{S_2}})$ and $ L_S= L_{S_1}+L_{S_2}$.  For equal cross sections  ${\mathcal A}_{S_1} = {\mathcal A}_{S_2}$ the parameter $R_n^a$ coincides with the normal state resistance of our junction $R_n$ and the result (\ref{IJeq}) immediately reduces to that of Ref. \cite{KDZ20}, cf. also Ref. \cite{ZZh}.

The last term $I_S^{\mathrm{ne}}$ in Eq. (\ref{IS}) represents an extra non-equilibrium contribution to the supercurrent which differs from zero provided $T_1 \neq T_2$. In the limit $T_{1,2} \gg E_{\mathrm{Th}}$ this term reduces to
\begin{equation}
I_S^{\mathrm{ne}} = \dfrac{1}{e}
\left(\dfrac{1}{T_1} - \dfrac{1}{T_2}\right) K(\chi),
\label{Ine}
\end{equation}
where
\begin{multline}
K(\chi)=
-
\dfrac{1}{4}
\int
\mathcal{G}
\dfrac{(G^L_{N_1} G^n_{N_2} - G^L_{N_2} G^n_{N_1})}{(G^L_{N_1} + G^L_{N_2})(G^n_{N_1} +G^n_{N_2})}
\varepsilon d \varepsilon
\\+
\dfrac{1}{4}
\int
\Biggl[
\dfrac{G_{S_1} G_{S_2}^n - G_{S_2} G_{S_1}^n}{G^n_{S_1} + G^n_{S_2}}
+
\mathcal{G}
\dfrac{G^{\mathcal{Y}}_{N_1} + G^{\mathcal{Y}}_{N_2}}{G^L_{N_1} + G^L_{N_2}}
\Biggr]
\dfrac{[G^L_{N_1}G^{\mathcal{Y}}_{N_2} - G^{\mathcal{Y}}_{N_1} G^L_{N_2}]
\varepsilon d \varepsilon}{
\det|\hat G_{S_1} + \hat G_{S_2} + \hat G_{N_1} + \hat G_{N_2}|}.
\label{Kchi}
\end{multline}
\begin{figure}
\begin{center}
\includegraphics[width=60mm]{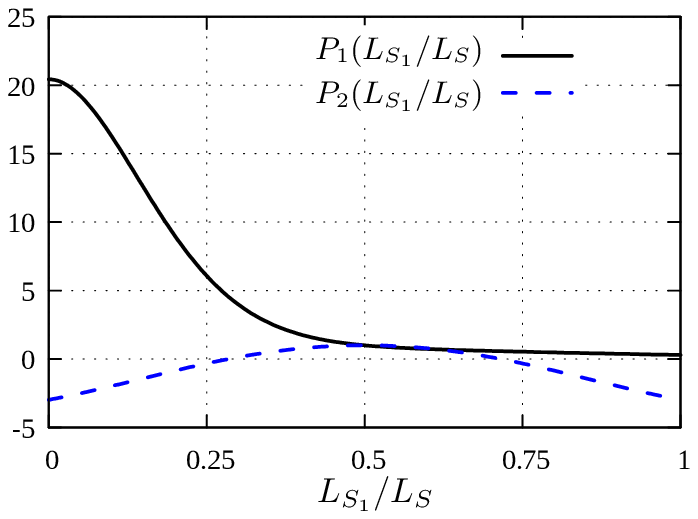}
\includegraphics[width=60mm]{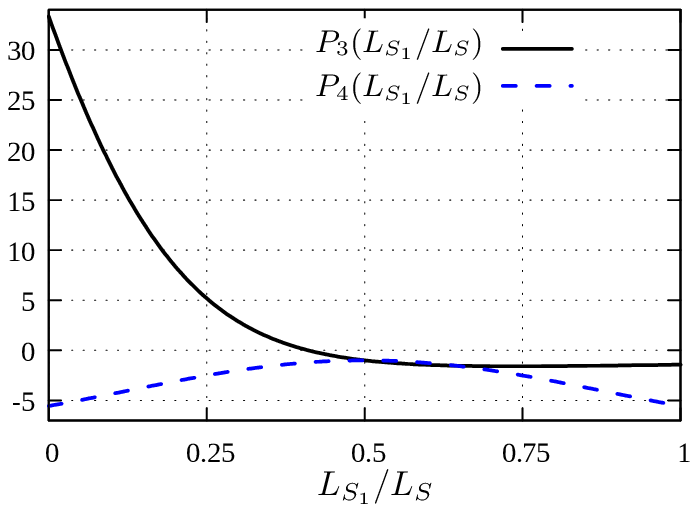}
\end{center}
\caption{Universal functions $P_{1,2}(L_{S_1}/L_S)$ (left panel) and $P_{3,4}(L_{S_1}/L_S)$ (right panel). 
We observe that $P_1(1/2) = P_2(1/2) = 1$, while $P_3(1/2) = P_4(1/2) = -1$.}
\label{P-fig}
\end{figure}

Our detailed numerical analysis of Eq. (\ref{Kchi}) demonstrates that the effect of electron-hole asymmetry on the function $K(\chi)$, though exists, always remains very small and, hence, can be safely neglected for {\it any} geometry of our $X$-junction. This is an important conclusion which allows to drop the conductance $G^{\mathcal{Y}}$ from Eq. \eqref{Kchi} and significantly simplify our further calculations. Setting $G^{\mathcal{Y}}$ to zero, from Eq. \eqref{Kchi} we obtain 
\begin{equation}
K(\chi)=
-
\dfrac{1}{4}
\int
\mathcal{G}(\varepsilon)W(\varepsilon)
\varepsilon d \varepsilon,
\quad
W(\varepsilon) = \dfrac{(G^L_{N_1} G^n_{N_2} - G^L_{N_2} G^n_{N_1})}{(G^n_{N_1} +G^n_{N_2})(G^L_{N_1} +G^L_{N_2})},
\label{Kchi2}
\end{equation}
Proceeding  along the lines with Ref. \cite{KDZ20} one can directly evaluate the functions $\mathcal{G}$ and $W$ in the high energy limit. Extrapolating this high energy expansion to the whole energy interval, evaluating the integral in Eq. \eqref{Kchi2} and combining the result with 
Eq. (\ref{Ine}), we get
\begin{multline}
I_S^{\mathrm{ne}} = 
\dfrac{4\varkappa^3}{(3+2\sqrt{2})^2} \dfrac{1101}{1250}
r_{N_1} r_{N_2}
\left(\dfrac{1}{T_1} - \dfrac{1}{T_2}\right)
\left(
\dfrac{L_S}{L_{N_2}} - \dfrac{L_S}{L_{N_1}}
\right)
\dfrac{ E_{\mathrm{Th}}}{eR_n^a}
\\\times
\left\{
\dfrac{\mathcal{A}_{S_1}}{2 \mathcal{A}_{S_2}} P_1(L_{S_1}/L)
+
\dfrac{\mathcal{A}_{S_2}}{2 \mathcal{A}_{S_1}}  P_1(L_{S_2}/L)
+
P_2(L_{S_1}/L)
\cos \chi
\right\}
\sin \chi ,
\label{Ktheor}
\end{multline}
where the universal functions $P_1$ and $P_2$ are displayed in Fig. \ref{P-fig} (left panel).

\begin{figure}
\begin{center}
\includegraphics[width=60mm]{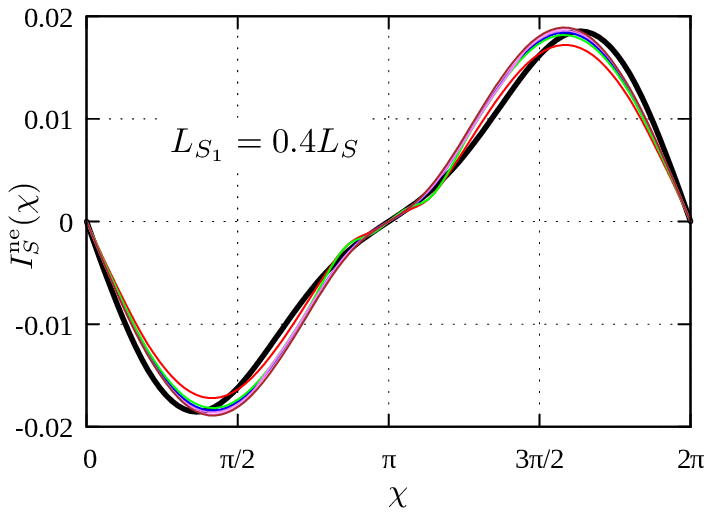}
\includegraphics[width=60mm]{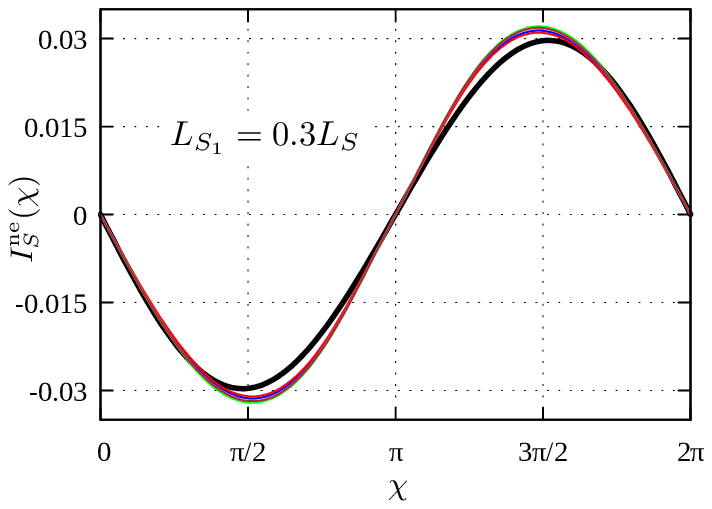}\\
\includegraphics[width=60mm]{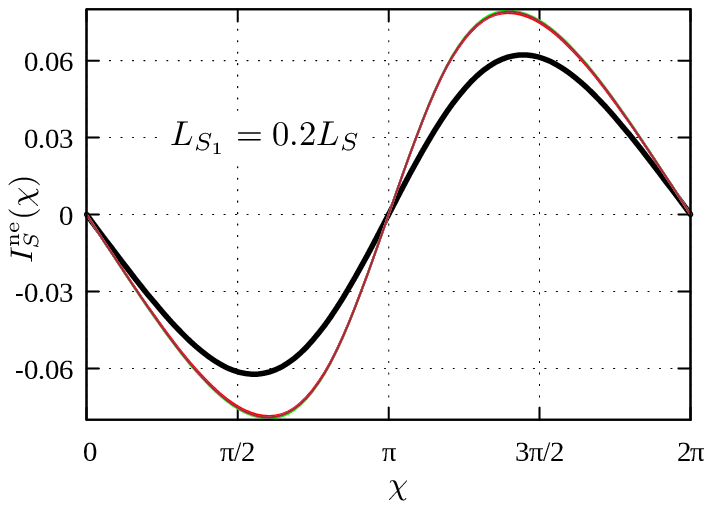}
\includegraphics[width=60mm]{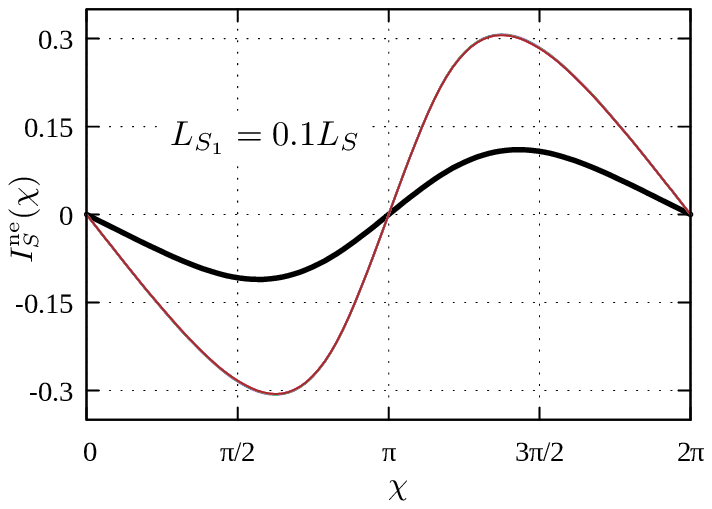}
\end{center}
\caption{Phase dependence of the non-equilibrium  term $I_S^{\mathrm{ne}}(\chi )$, normalized by the factor $(1/T_1-1/T_2)E^2_{\mathrm{Th}}/eR_n^a$. 
Thick black lines represent our analytical results, whereas numerically exact results are shown by thin color lines. 
Top left panel: The parameters $L_{S_1}=0.4L_S$ and $T_1=65 E_{\mathrm{Th}}$ apply to all curves, different thin color curves correspond to different temperatures $T_2$ ($40 E_{\mathrm{Th}}$, $50 E_{\mathrm{Th}}$, $60 E_{\mathrm{Th}}$, $70 E_{\mathrm{Th}}$, $80 E_{\mathrm{Th}}$, $90 E_{\mathrm{Th}}$).
Top right panel: The parameters $L_{S_1}=0.3L_S$ and $T_1=60 E_{\mathrm{Th}}$ apply to all curves, different thin color curves correspond to different temperatures $T_2$ ($35 E_{\mathrm{Th}}$, $45 E_{\mathrm{Th}}$, $55 E_{\mathrm{Th}}$, $65 E_{\mathrm{Th}}$, $75 E_{\mathrm{Th}}$, $85 E_{\mathrm{Th}}$).
Bottom left panel: The parameters $L_{S_1}=0.2L_S$ and$T_1=55 E_{\mathrm{Th}}$ apply to all curves, different thin color curves correspond to different temperatures $T_2$ ($30 E_{\mathrm{Th}}$, $40 E_{\mathrm{Th}}$, $50 E_{\mathrm{Th}}$, $60 E_{\mathrm{Th}}$, $70 E_{\mathrm{Th}}$, $80 E_{\mathrm{Th}}$).
Bottom right panel: The parameters $L_{S_1}=0.1L_S$ and $T_1=50 E_{\mathrm{Th}}$ apply to all curves, different thin color curves correspond to different temperatures $T_2$ ($25 E_{\mathrm{Th}}$, $35 E_{\mathrm{Th}}$, $45 E_{\mathrm{Th}}$, $55 E_{\mathrm{Th}}$, $65 E_{\mathrm{Th}}$, $75 E_{\mathrm{Th}}$). Other parameters are: $\Delta=1000 E_{\mathrm{Th}}$, $L_{N_1}=L_S$, $L_{N_2}=3 L_S$, $\mathcal{A}_{S_1} = \mathcal{A}_{S_2} = \mathcal{A}_{N_1} = \mathcal{A}_{N_2}$.}
\label{K-chi-LN1-1-LN2-3-fig}
\end{figure}

Equation \eqref{Ktheor} demonstrates that in a wide temperature interval $T_{1,2}>E_{\mathrm{Th}}$ the non-equilibrium contribution to the Josephson current  is described by a universal power law dependence $\propto 1/T_1 - 1/T_2$, whereas the phase dependence of $I_S^{\mathrm{ne}}$ depends on the junction geometry only. In Fig. \ref{K-chi-LN1-1-LN2-3-fig} we display the results of our numerical solution of the Usadel equation together with those defined by Eq. \eqref{Ktheor}.

For partially symmetric junction with $L_{S_1}=L_{S_2}$ and $\mathcal{A}_{S_1} = \mathcal{A}_{S_2}$ we have $P_1=P_2=1$ and the phase dependence of $I_S^{\mathrm{ne}}$ reduces to the peculiar form $\cos^2(\chi/2)\sin\chi$ \cite{KDZ20}. For strongly asymmetric junctions the function $P_1$ increases by about an order of magnitude, whereas $P_2$ varies slightly (see Fig. \ref{P-fig}). In this case the current-phase relation approaches a sin-like form. The results displayed in Fig. 3 demonstrate that our analytic formula \eqref{Ktheor} is in a good agreement with numerically exact results for $I_S^{\mathrm{ne}}$  as long as the lengths $L_{S_{1,2}}$ remain not very small $\min(L_{S_1}, L_{S_2}) \gtrsim 0.2 L_S$.

\section{Supercurrent stimulation and $\pi$-junction states}

The above results demonstrate that in the presence of a temperature gradient and at sufficiently high temperatures the supercurrent is not anymore exponentially small due to the non-equilibrium contribution (\ref{Ktheor}). In other words, at $T_{1,2}\gg E_{\mathrm{Th}}$ the Josephson current in our $X$-junction is stimulated by the temperature gradient. The relative magnitude of $I_S^{\mathrm{ne}}$ -- as compared to the equilibrium term $I_J$ (\ref{IJeq}) -- can further be enhanced by a proper choice of the junction geometry. For instance, in a strongly asymmetric limiting case  $L_{S_1} \ll L_{S_2}$ and $\mathcal{A}_{S_1} \gg \mathcal{A}_{S_2} + \mathcal{A}_{N_1}+ \mathcal{A}_{N_2}$ we obtain
\begin{equation}
I_S^{\mathrm{ne}} = 
\dfrac{32\varkappa}{(3+2\sqrt{2})^2} \dfrac{1101}{1250}
r_{N_1} r_{N_2}
\left(\dfrac{1}{T_1} - \dfrac{1}{T_2}\right)
\left(
\dfrac{L_S}{L_{N_2}} - \dfrac{L_S}{L_{N_1}}
\right)
\dfrac{E_{\mathrm{Th}}}{eR_n^a}
P_1(L_{S_1}/L)
\sin \chi .
\label{Ktheoropt}
\end{equation}

\begin{figure}
\begin{center}
\includegraphics[width=60mm]{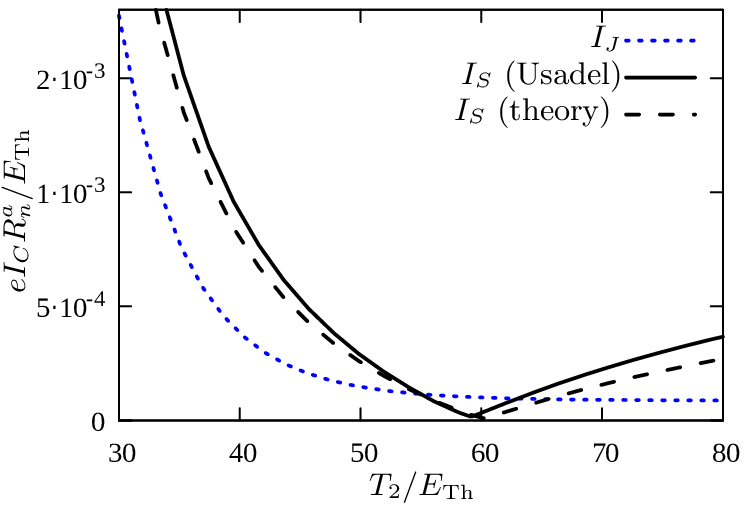}
\includegraphics[width=60mm]{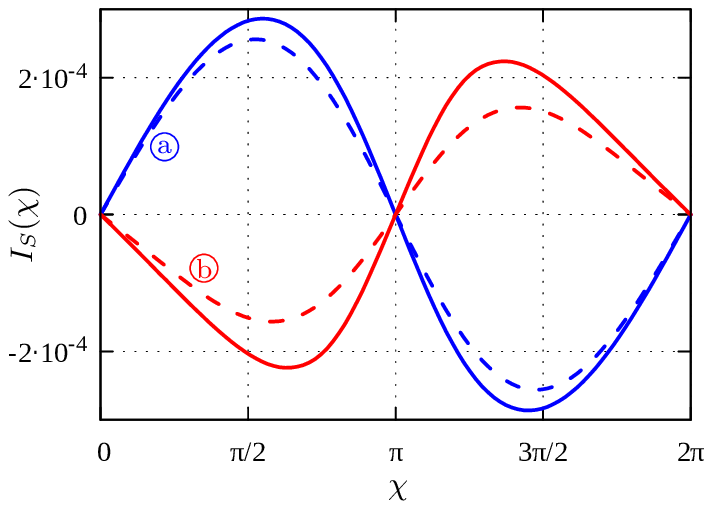}
\end{center}
\caption{Left panel: Josephson critical current $I_C\equiv \max |I_S|$ as a function of $T_2$. Solid lines correspond to our numerically exact solution, dashed lines indicate the result \eqref{IS} combined with \eqref{Ine} and \eqref{Ktheor}, dotted line represent the quasi-equilibrium contribution $r_{N_2} I_J(T_1,\pi/2) + r_{N_1} I_J(T_2,\pi/2)$ to $I_S$. 
Right panel: The phase dependencies of the Josephson current $I_S(\chi)$ for $T_2 = 50 E_{\mathrm{Th}}$ (``a'' curves) and $T_2 = 70 E_{\mathrm{Th}}$ (``b'' curves). Other parameters are the same for both panels: $T_1=55 E_{\mathrm{Th}}$, $L_{N_1}=L_S$, $L_{N_2}=3 L_S$, $L_{S_1}=0.2L_S$ and $\mathcal{A}_{S_1} = \mathcal{A}_{S_2} = \mathcal{A}_{N_1} = \mathcal{A}_{N_2}$.}
\label{ISc-LS1-2-LS2-8-LN1-1-LN2-3-fig}
\end{figure}

\begin{figure}
\begin{center}
\includegraphics[width=60mm]{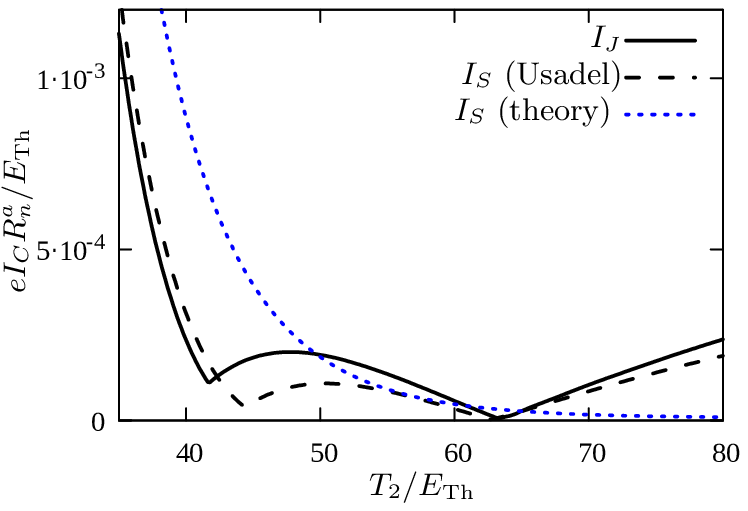}
\includegraphics[width=60mm]{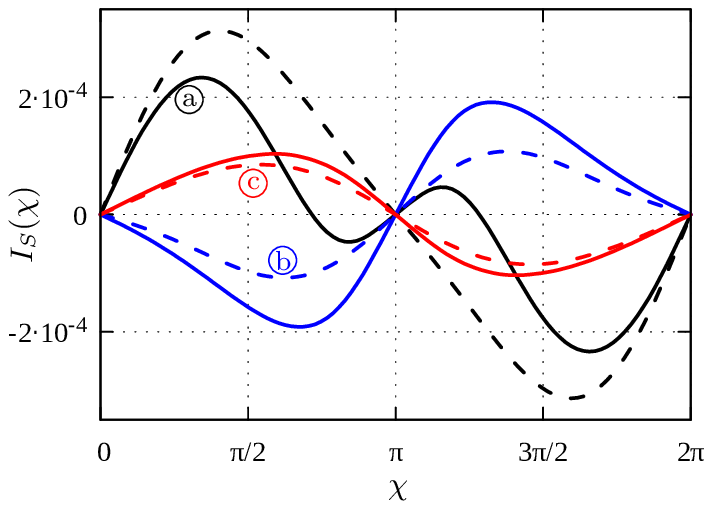}
\end{center}
\caption{Left panel: The same as in Fig. 4 (left panel).
Right panel:  The phase dependencies of the Josephson current $I_S(\chi)$ for $T_2 = 40 E_{\mathrm{Th}}$ (``a'' curves), $T_2 = 50 E_{\mathrm{Th}}$ (``b'' curves), and $T_2 = 70 E_{\mathrm{Th}}$ (``c'' curves). Other parameters are the same for both panels: $T_1=65 E_{\mathrm{Th}}$, $L_{N_1}=3 L_S$, $L_{N_2}=L_S$, $L_{S_1}=0.2L_S$ and $\mathcal{A}_{S_1} = \mathcal{A}_{S_2} = \mathcal{A}_{N_1} = \mathcal{A}_{N_2}$.}
\label{ISc-LS1-2-LS2-8-LN1-3-LN2-1-fig}
\end{figure}

It is easy to verify that this term dominates the supercurrent already at  $\min T_{1,2} \gtrsim 30E_{\mathrm{Th}}$, i.e. at considerably
lower temperatures than in symmetric junctions with $L_{S_1} = L_{S_2}$ and $\mathcal{A}_{S_1} = \mathcal{A}_{S_2} = \mathcal{A}_{N_{1}} = \mathcal{A}_{N_{2}}$, where the analogous condition reads $T_{1,2} \gtrsim 70 E_{\mathrm{Th}}$, cf. Ref. \cite{KDZ20}. We also observe from Figs. \ref{ISc-LS1-2-LS2-8-LN1-1-LN2-3-fig} and \ref{ISc-LS1-2-LS2-8-LN1-3-LN2-1-fig} below that for partially asymmetric junctions with  $L_{S_1} \ll L_{S_2}$ and $\mathcal{A}_{S_1} = \mathcal{A}_{S_2} = \mathcal{A}_{N_{1}} = \mathcal{A}_{N_{2}}$ nonequilibrium effects become visible only at $T_{1,2}/E_{\mathrm{Th}}\gtrsim 40\div 50 $. Thus, we conclude that strongly asymmetric $X$-junctions with $L_{S_1} \ll L_{S_2}$ and $\mathcal{A}_{S_1} \gg \mathcal{A}_{S_2} + \mathcal{A}_{N_1}+ \mathcal{A}_{N_2}$ appear to be most suitable candidates for observing the non-equilibrium Josephson current  defined in Eq. (\ref{Ktheoropt}).
 
Our general analysis also applies provided the temperature difference becomes large. In order to address this particular situation let us set $T_1 \to 0$ and $T_2\gg E_{\mathrm{Th}}$.  As before, the effect of electron-hole asymmetry on $I_S$ remains weak and can be neglected in this case. Then Eq.  (\ref{IS}) reduces to the form (cf. also Ref. \cite{KDZ20}):
\begin{equation}
I_S = r_{N_2}^n I_J(0,\chi)+I^{\mathrm{ne}}_S,\quad    I^{\mathrm{ne}}_S=-
\dfrac{1}{2e}
\int
\mathcal{G}(\varepsilon)W(\varepsilon)
\sgn \varepsilon
 d \varepsilon.
\end{equation}
In order to obtain a simple analytic estimate for the equilibrium Josephson current at zero temperature $I_J(0,\chi)$ we again
employ the high energy solution of the Usadel equation extending it to all energies. Then from Eq. (\ref{IJ}) we get
\begin{equation}
I_J(0,\chi)
=
\dfrac{32\varkappa }{3+2\sqrt{2}}\dfrac{E_{\mathrm{Th}} }{e R_n^a}
 \sin\chi .
 \label{IJ0}
\end{equation}
Remarkably, in the case of SNS junctions (with $\varkappa=2$ and $R_n^a\equiv R_n$) this simple estimate provides a very
accurate value of the critical current $I_C(0)$ (cf., Ref. \cite{GreKa}). As expected, it cannot capture the non-sinusoidal character
of the current-phase relation in this case, which is, however, unimportant for our present purposes.

Evaluating the non-equilibrium term $I^{\mathrm{ne}}_S$ we employ the same approximation combined with the expression for $W(\varepsilon)$ (\ref{Kchi2}) and find
\begin{multline}
I^{\mathrm{ne}}_S =\dfrac{0.72\varkappa^3  }{(3+2\sqrt{2})^2}r_{N_1}r_{N_2}
\left(
\dfrac{L_S}{L_{N_2}} - \dfrac{L_S}{L_{N_1}}
\right)
\dfrac{E_{\mathrm{Th}} }{ e R_n^a}\Biggl\{
\dfrac{\mathcal{A}_{S_1}}{\mathcal{A}_{S_2}} P_3(L_{S_1}/L)
+\\+
\dfrac{\mathcal{A}_{S_2}}{\mathcal{A}_{S_1}} P_3(L_{S_2}/L)
+
2 P_4( L_{S_1}/L) \cos \chi
\Biggr\} \sin\chi,
\end{multline}
where the functions $P_3$ and $P_4$ are displayed in Fig. 2 (right panel).

In particular, in strongly asymmetric $X$-junctions with $L_{S_1} \ll L_{S_2}$ and $\mathcal{A}_{S_1} \gg \mathcal{A}_{S_2} + \mathcal{A}_{N_{1}} + \mathcal{A}_{N_{2}}$ this result reduces to
\begin{equation}
I^{\mathrm{ne}}_S =
\dfrac{11.52\varkappa  }{(3+2\sqrt{2})^2}r_{N_1}r_{N_2}
\left(
\dfrac{L_S}{L_{N_2}} - \dfrac{L_S}{L_{N_1}}
\right)\dfrac{E_{\mathrm{Th}} }{ e R_n^a} P_3(L_{S_1}/L)
 \sin\chi .
\end{equation}
We observe that for large temperature gradients the magnitude of the non-equilibrium contribution to the Josephson current 
$I_S^{\mathrm{ne}}$ -- apart from some geometry factors -- can be of the same order as that of the equilibrium supercurrent $I_J$  (\ref{IJ0}).
These two contributions to $I_S$ can have either the same or opposite signs depending on whether $L_{N_1}$ is longer or shorter than $L_{N_2}$.

In order to analyze the temperature dependence of the critical Josephson current $I_C=\max I_S(\chi)$ we will again employ Eq. (\ref{IS}). At the temperatures exceeding the Thouless energy we have $I_J \propto \sin \chi$ with a positive prefactor (see Eq. \eqref{IJeq}), whereas the phase dependence of the non-equilibrium term $I_S^{\mathrm{ne}}$ \eqref{Ktheor} is described by a somewhat distorted $\sin \chi$ function with either positive or negative sign in front of it depending on the sign of the product $(T_2 - T_1) (L_{N_1}-L_{N_2})$. 

Keeping one of the two temperatures (e.g., $T_1$) fixed and varying $T_2$ we observe that the junction behavior strongly depends on the relation between the lengths $L_{N_1}$ and $L_{N_2}$. For $L_{N_1}<L_{N_2}$ the sign of  $I_S^{\mathrm{ne}}$ is remains positive at $T_2 < T_1$ 
and turns negative at $T_2 > T_1$, implying that at sufficiently high temperatures this term dominates over exponentially decaying contributions containing $I_J$ \eqref{IJeq} and, hence, our $X$-junction switches to the $\pi$-junction state. This behavior is illustrated in Fig. \ref{ISc-LS1-2-LS2-8-LN1-1-LN2-3-fig} (left panel). In the right panel of Fig. \ref{ISc-LS1-2-LS2-8-LN1-1-LN2-3-fig} we display typical current-phase dependencies corresponding to both $0$- and $\pi$-junction states realized in our structure. 

For $L_{N_1} > L_{N_2}$ our $X$-junction may already exhibit two transitions between $0$- and $\pi$-junction states. In  this case our system remains in the 0-junction state provided $T_2$ remains low enough to keep the quasi-equilibrium term larger than  $I_S^{\mathrm{ne}}$. However, since with increasing $T_2$ (albeit still $T_2< T_1$)  the contribution $\propto I_J$ decays much faster as the non-equilibrium one (now having a negative sign), the $X$-junction eventually switches to the $\pi$-junction state. Increasing $T_2$ further we reach the point  $T_2=T_1$ where  $I_S^{\mathrm{ne}}$ changes its sign, thus signaling the transition back to the $0$-junction state at $T_2$ slightly below $T_1$. This behavior is illustrated in Fig. \ref{ISc-LS1-2-LS2-8-LN1-3-LN2-1-fig}.

In summary, we evaluated the Josephson current $I_S$ across an $X$-junction exposed to a temperature gradient and demonstrated that the effect of supercurrent stimulation can be further enhanced by a proper choice of geometric parameters for our device. We also predicted a non-trivial power-law temperature dependence of $I_S$ and showed that our $X$-junction may exhibit transitions between 0- and $\pi$-junction states controlled both by the temperature gradient and the circuit topology. Note that the junction topology can further be modified to embrace, e.g., the one addressed in Ref. \cite{VH} in the context of the thermoelectric effect. The whole analysis (to be published elsewhere \cite{KZ20}) becomes much more involved but some of our key observations remain applicable also in this case. It would be interesting to test our predictions in future experiments.

This work was supported in part by RFBR grant No. 18-02-00586.


\begin{thebibliography}{99}
\bibitem{Jos} B.D. Josephson, Phys. Lett. \textbf{1}, 251 (1962)
\bibitem{ZZh} A.D.  Zaikin, G.F.  Zharkov,  Fiz.  Nizk.  Temp. \textbf{7},  375 (1981) [Sov. J. Low Temp. Phys. \textbf{7}, 181 (1981)]
\bibitem{GreKa} P. Dubos, H. Courtois, B. Pannetier, F.K. Wilhelm, A.D. Zaikin, G. Sch\"on, Phys. Rev. B \textbf{63},  064502 (2001)
\bibitem{BWBSZ1999} W. Belzig, F.K. Wilhelm, C. Bruder, G. Sch{\"o}n, A.D. Zaikin, Superlatt. Microstruct. \textbf{25}, 1251 (1999)
\bibitem{GKI} A.A. Golubov, M.Yu. Kupriyanov, E. Il'ichev, Rev. Mod. Phys. \textbf{76}, 411 (2004)
\bibitem{Aslamazov82} L.G. Aslamazov, S.V. Lempitskii, Zh. Eksp. Teor. Fiz. \textbf{82}, 1671 (1982) 
[Sov. Phys. JETP \textbf{55}, 967 (1982)]
\bibitem{Zaikin83} A.D. Zaikin, Zh. Eksp. Teor. Fiz. \textbf{84}, 1560 (1983) [Sov. Phys. JETP \textbf{57}, 910 (1983)]
\bibitem{Volkov} A.F. Volkov, Phys. Rev. Lett. \textbf{74}, 4730 (1995)
\bibitem{WSZ} F.K. Wilhelm, G. Sch\"on, A.D. Zaikin, Phys. Rev. Lett. \textbf{81}, 1682 (1998)
\bibitem{Yip} S. Yip, Phys. Rev. B \textbf{58}, 5803 (1998)
\bibitem{Teun} J.J.A. Baselmans, A.F. Morpurgo, B.J. van Wees, T.M. Klapwijk, Nature \textbf{397}, 43 (1999)
\bibitem{KDZ20} M.S. Kalenkov, P.E. Dolgirev, A.D. Zaikin, Phys. Rev. B \textbf{101}, 180505(R) (2020)
\bibitem{Gre} H. Courtois, P. Gandit, D. Mailly, B. Pannetier, Phys. Rev. Lett. \textbf{76}, 130 (1996)
\bibitem{GWZ} A.A. Golubov, F.K. Wilhelm, A.D. Zaikin, Phys. Rev. B \textbf{55}, 1123 (1997)
\bibitem{Ginzburg} V.L. Ginzburg, Rev. Mod. Phys. \textbf{76}, 981 (2004)
\bibitem{VH} P. Virtanen, T.T. Heikkil\"a, Phys. Rev. Lett. \textbf{92}, 177004 (2004)
\bibitem{KZ20} M.S. Kalenkov, A.D. Zaikin, in preparation
\end{thebibliography}
\end{document}